\documentclass[12pt,a4paper]{article}

\usepackage{color}
  \usepackage{a4wide}
  \usepackage{latexsym}
  \usepackage{epsf}
  \usepackage{amssymb,bbm}
  \usepackage{graphicx}
  \usepackage{amsmath, cite}
  \usepackage{amsmath,amssymb,amsthm}
  \usepackage{verbatim}
  \usepackage{hyperref}
  \usepackage{amsfonts}
  \usepackage{xfrac}
  \usepackage{slashed}
  \usepackage{tikz}
  \usepackage{cancel}
	\usepackage{verbatim}
  
\renewcommand{\d}{\textrm{d}}
\newcommand{\e}{\textrm{e}}

\newcommand{\w}{\wedge}

\newcommand{\be}{\begin{equation}}
\newcommand{\ee}{\end{equation}}

\newcommand{\roughly}[1]{\mathrel{\raise.3ex\hbox{$#1$\kern-0.85em
\lower1ex\hbox{$\sim$}}}}

\def\be{\begin{equation}}
\def\ee{\end{equation}}
\def\bea{\begin{eqnarray}}
\def\eea{\end{eqnarray}}
\def\ba{\begin{align}}
\def\ea{\end{align}}
\def\bga{\begin{gather}}
\def\ega{\end{gather}}

\def\pd{\partial}

\def\pref#1{(\ref{#1})}

\def\cQ{{\cal Q}}

\def\dsl{\hbox{/\kern-.5300em$\partial$}}

\newcommand{\bmat}{\left(\begin{array}}
\newcommand{\emat}{\end{array}\right)}

\def\s2{\frac{1}{2}}

\def\exd{{\rm d}}

\pdfoutput=1

\begin{document}
\numberwithin{equation}{section}
\begin{flushright}
\small IPhT-t15/175\\
\normalsize
\end{flushright}

\begin{center}

\vspace{1.0 cm}

{\LARGE {\bf Domain walls inside localised orientifolds }}  \\

\vspace{1.5 cm} {\large  J. Bl{\aa }b\"ack$^a$, E. van der Woerd$^b$,\\ \vspace{0.2 cm}  T. Van
Riet$^b$ and M.~Williams$^b$}\\
\vspace{1.5 cm}{$^{a }$Institut de Physique Th\'eorique, CEA
Saclay, \\CNRS URA 2306 ,  F-91191 Gif-sur-Yvette, France}\\
\vspace{0.2 cm}  \vspace{.15 cm} {$^b$ Instituut voor Theoretische Fysica, K.U. Leuven,\\
Celestijnenlaan 200D B-3001 Leuven, Belgium}

\vspace{0.7cm} {\small \upshape\ttfamily johan.blaback @ cea.fr,  ellen, thomasvr, m.williams @ itf.fys.kuleuven.be }  \\

\vspace{1.2cm}

{\bf Abstract}
\end{center}

{\small The equations of motion of toroidal orientifold compactifications with fluxes are in one-to-one correspondence with gauged supergravity if the orientifold (and D-brane) sources are smeared over the compact space. This smeared limit is identical to the approximation that ignores warping.
 It is therefore relevant to compare quantities obtained from the gauged supergravity with the true 10d solution with localised sources. In this paper we find the correspondence between BPS domain walls in gauged SUGRA and 10D SUGRA with localised sources. 
Our model is the simplest orientifold with fluxes we are aware of: an O6/D6 compactification on $\mathbb{T}^3/\mathbb{Z}_2$ in massive IIA with $H_3$-flux. The BPS domain walls correspond to a O6/D6/NS5/D8 bound state. Our analysis reveals that the domain wall energy computed in gauged SUGRA is unaffected by the localisation of the O6/D6 sources. }

\setcounter{tocdepth}{2}
\newpage
\tableofcontents

\section{Introduction}
Flux compactifications of 10-dimensional supergravity  often invoke orientifold and D-brane sources for good reasons: 1) the sources break part of the supersymmetry; 2) they are necessary for achieving a hierarchical separation between the Kaluza-Klein scale and the vacuum energy; \cite{DeWolfe:2005uu, Caviezel:2008ik,  Tsimpis:2012tu, Petrini:2013ika} and 3) they are an essential ingredient for constructing dS solutions \cite{Maldacena:2000mw}. For the latter two effects to take place one needs to make sure that there is a net negative orientifold tension left over. This means that the RR tadpole for the orientifold should not be canceled with D-branes but with fluxes \cite{Dasgupta:1999ss, Giddings:2001yu}. In any case, the presence of orientifolds and D-branes causes the compactification to be  \emph{warped}\footnote{Warping would be absent when the orientifold charge is cancelled by D-brane charges that are right on top of the orientifold.}. Warping is usually associated with a non-vanishing function $e^{2A(x, z)}$ in front of the four-dimensional metric
\be
\d s^2_{10} =  e^{2A(x, z)} g^{(4)}_{\mu\nu}\d x^{\mu}\d x^{\nu} + g^{(6)}_{ij}(z)\d z^i\d z^j\,. 
\ee 
But in essence \emph{warping} should be defined as the  collection of all fields $\Psi$ that acquire a non-trivial dependence on the extra-dimensional coordinates through the delta-functions that represent the brane sources. So 
this can also include the dilaton and RR form fields to which the brane couples. Symbolically one can write
\be\label{BOX}
\Box_6 \Psi = \text{fluxes}  + \delta\,,
\ee
where ``fluxes'' denotes the finite terms that are typically some combination of fluxes. If the sources are \emph{smeared} then $\delta$ is replaced by a finite form such that the right hand side of (\ref{BOX}) vanishes and the field $\Psi$ does not have dependence on internal coordinates. This is for instance how one can think of IIB orientifolds with 3-form fluxes \cite{Giddings:2001yu} as Calabi--Yau compactifications; in the smeared limit the backreaction of the orientifolds exactly cancels the backreaction of the 3-form fluxes such that the Calabi-Yau geometry solves the 10D equations of motion (EOM) \cite{Grana:2006kf, Blaback:2010sj, Andriot:2015sia}. The same holds for AdS vacua in massive IIA \cite{Acharya:2006ne, Saracco:2012wc}.

Smearing naively looks like a radical approximation, but it makes sense from the point of view of effective field theory. The delta-function lives in the compact space, but if one coarse-grains (integrates) over distances smaller than the KK scale the delta-function can be replaced by any function that has the same integral. This is manifested by the fact that the smearing procedure often allows a consistent truncation to a lower-dimensional supergravity\footnote{In \cite{Danielsson:2013qfa} a 1/2 BPS AdS vacuum of type IIA SUGRA was found not to allow a gauged SUGRA description upon smearing the sources. This is consistent with the fact that the latter vacua have no separation of scales and the effective field theory should be 10-dimensional. Although when enough modes are truncated a gauged SUGRA does arise \cite{Passias:2015gya}.}.

The main motivation of our work is to understand the extent to which the smeared approximation is accurate and to identify which quantities get corrected due to localised sources. In particular, we are motivated by seemingly contradictory statements in the literature. On the one hand it is believed that smearing is a good approximation when the space over which the branes are smeared is small (in string units) and, on the other hand, it is believed that exactly the large volume limit justifies ignoring warping, see e.g. \cite{Douglas:2007tu}.
 But we have argued that ignoring warping is mathematically the same as smearing the sources. Hence we arrive at what seems as a contradiction. We have not (yet) resolved this puzzle and we hope to come back to this in the near future.

One can think of the dependence of fields $\Psi$ on internal coordinates as Kaluza-Klein modes and hence they are typically discarded. The smearing can then be understood as a Fourier-expansion of the delta-function where only the constant term is kept. There have been some discussions in the literature on how these KK modes should be integrated out and this goes by the name of \emph{warped effective field theory} (WEFT) and some essential references on this topic are \cite{Giddings:2005ff, Shiu:2008ry, Frey:2008xw, Martucci:2009sf, Underwood:2010pm, Frey:2013bha, Martucci:2014ska, Grimm:2015mua}. In the smeared limit the coordinate dependences are simply eliminated and not integrated over, whereas in WEFT there is a non-trivial integral.

Instead of WEFT we aim at finding an \emph{exact} correspondence between essential data of the lower-dimensional supergravity and the 10-dimensional theory. In particular we focus on solitonic BPS states of the lower-dimensional supergravity that are domain walls (co-dimension one objects). Domain walls are supported by the scalar fields, such as the dilaton, and are therefore sensitive to the details of the scalar kinetic terms and the scalar potential. The scalar sector is crucial for understanding the vacuum structure of the lower-dimensional theory and our interest in the BPS domain walls lies in the fact that they probe this sector. Apart from being part of the non-perturbative spectrum of states, domain walls are useful for describing instanton transitions between different flux vacua \cite{Ceresole:2006iq}\footnote{ Interestingly, it has been claimed \cite{Ahlqvist:2010ki, Ahlqvist:2012za, Masoumi:2011kd}
that the warping corrections to the unwarped effective field theory are relevant effects for the tunneling process.}.

 Domain walls are characterised by their energy (and tension). The main goal of this paper is to understand whether or not the energy is affected by smearing the background orientifolds/D-branes that sustain the compactification. The latter requires a full 10-dimensional treatment in which the domain walls can be understood as certain branes wrapping internal cycles such that they have a single co-dimension inside the non-compact part of spacetime. For clarity we emphasize that at all times the domain walls themselves are localised objects, but it is the orientifolds and D-branes that sustain the compactification which we consider both smeared and localised.
 
  We perform this computation in a simple flux compactification such that everything is fully explicit and computable. In the end we demonstrate that the energy is not altered by localisation because the energy can be used to identify the domain wall solutions inside the infinite set of solutions to the 1/4 BPS equations in 10D. 

The rest of this paper is organised as follows. In section \ref{background} we present the simple explicit flux compactification:
massive IIA on $\mathbb{T}^3/\mathbb{Z}_2$ with space-filling O6/D6 sources and a combination of NSNS 3-form flux $H_3$ and RR Romans mass $F_0$ to cancel the tadpole. The compactification with smeared sources leads to a  specific half-maximal gauged SUGRA in 7D whose vacuum is  Minkowski and breaks all supersymmetries. The same vacuum persists when the sources are localised.  In section \ref{SUGRASOL} we first construct the domain wall solutions of the 7D gauged SUGRA, which lift to 10D solutions with smeared O6/D6 sources. Then we construct the domain wall solutions in case the O6/D6 sources are localised. These walls correspond to D8 branes wrapping the 3-torus and NS5 branes inside the non-compact space. In section \ref{TENSION} we then compute the wall energy in the smeared limit and in the localised limit. The latter expression is then analysed in detail and we come to the conclusion that the two expressions match exactly. We end with a discussion in section \ref{Discussion}. We  added an appendix in which we analyse in some detail the Minkowski vacuum of the orientifold compactification by describing the solutions to the generalised Laplace equation for the warpfactor.

\section{The background}\label{background}

\subsection{The $\mathbb{T}^3/\mathbb{Z}_2$ orientifold in massive IIA}
We consider massive IIA supergravity \cite{Romans:1985tz}. In our conventions\footnote{Herein, we take the 10D gravitational constant, $\kappa^2$, equal to unity.} the action in 10-dimensional Einstein frame is
\begin{align}
\mathcal{S} & = \int \star_{10} \Bigl(\mathcal{R} - \tfrac{1}{2}(\partial\phi)^2  - \tfrac{1}{2\, 3!}\e^{-\phi}H_3^2  - \tfrac{1}{2}\e^{\tfrac{5}{2}\phi}F_0^2 - \tfrac{1}{2\, 2!}\e^{\tfrac{3}{2}\phi}F_2^2 - \tfrac{1}{2\, 4!}\e^{\tfrac{1}{2}\phi}F_4^2\Bigr)\nonumber \\
& + \int \frac{m^2}{40}B_2^5 +\frac{m}{6}B_2^3\wedge \d C_3 +\frac{1}{2}\d C_3\wedge \d C_3\wedge B_2 \,.
\end{align}
The square of a $p$-form $F^2$ is defined as $F_{ab\ldots}F^{ab\ldots}$. The various field strengths that appear in the action are
\begin{align}
& H_3=\d B_2 \,,\nonumber\\
& F_2 =  \d C_1 + F_0B_2\,,\nonumber\\
& F_4 = \d C_3 - H_3\wedge C_1 + \tfrac{1}{2} F_0 B_2\wedge B_2\,.
\end{align}

As a background we take $\mathcal{M}_7\times \mathbb{T}^3/\mathbb{Z}_2$ with space-filling O6 sources that sit at the 8 fixed points of the $\mathbb{Z}_2$ involution:
\begin{equation}
(a, b, c)\,,\qquad \text{where}\qquad a, b, c \in \{0, 1/2 \}\,.
\end{equation} 
To cancel the RR tadpole 
\be\label{tadpole}
\int F_0 H_3 - Q_{D6} =0 \,,
\ee
we need non-zero Romans mass $F_0=m$ and $H_3$-flux filling the compact dimensions. The various vacua in 7 dimensions differ by the number $N$ of D6 branes in the compact manifold. If we denote the flux quanta of $F_0$ and $H_3$ by the integers $M$ and $n$, the allowed values for $N$ are obtained from the tadpole condition (\ref{tadpole}) to be: 
\begin{equation}\label{quant1}
nM = 16 - N \ .
\end{equation}
The corresponding gravity solution of this orientifold vacuum was first discussed in \cite{Blaback:2010sj} (see also \cite{Junghans:2013xza}) and in the appendix we provide a more in-depth study.

\subsection{Half maximal gauged supergravity in $D=7$}

 From the parity rules for O6 planes ($B_2$ and $C_1$ are odd, $C_3$ is even) we derive the following bosonic field content in 7 dimensions: the metric field, $10  + 3N$  scalars, $6+ N$ vectors and a three-form. As explained in the appendix of \cite{Danielsson:2013qfa} the effective theory in $D=7$ is a half-maximal gauged supergravity  coupled to $N$ vector multiplets with the following scalar coset
\begin{equation}
\mathbb{R}^{+}\times\frac{\textrm{SO}(3, 3+N)}{\textrm{SO}(3)\times \textrm{SO}(3+N)}\, .
\end{equation}
The equations of motion of this gauged supergravity lift to the equations of motion of massive IIA supergravity with \emph{smeared} O6/D6 planes \cite{Blaback:2013taa}.

For the purpose of constructing domain wall solutions in the next section,  we only need a truncation of this theory to the 7-dimensional metric and two real scalar fields, which was described in \cite{Blaback:2012mu}. For the full dimensional reduction of the bosonic sector, we refer to the appendix of \cite{Danielsson:2013qfa} or \cite{Dibitetto:2015bia} for a general treatment of  half-maximal supergravity in $D=7$.

The Ansatz for the consistent reduction is given by
\begin{align}
&  \d s^2_{10} = \e^{2\alpha\varphi} \d s_7^2 + \e^{2\beta\varphi}\delta_{ij}\d y^i\d y^j\,, \nonumber\\
& H_3= h\,\d \theta^1\wedge\d \theta^2\wedge\d \theta^3\,, \nonumber\\
&  F_0 = m\,,\label{oxidation}
\end{align}
in the Einstein frame, with 
\begin{equation}
\alpha=\tfrac{1}{4}\sqrt{\tfrac{3}{5}}\,,\qquad \beta=-5\alpha/3\,,
\end{equation}
and $h$ is the $H_3$-flux quantum. Tadpole cancellation requires us to take the tension of the O6 plane to be\footnote{This also corrects a typo in \cite{Blaback:2013taa}, in which the tadpole is said to imply $T=-2hm$, in equation (3.7).}
\begin{equation}
\kappa_7^2\,T_{6} = - h m 
\end{equation}
where $1/\kappa_7^{2}:= V_3/\kappa^2 = V_3$ is the volume of the transverse space spanned by the $\exd y^i$'s.
Our conventions are such that $h$ and $m$ are both positive. The only contribution to the scalar potential in seven dimensions comes from the Romans mass, the $H_3$-flux, and the orientifold tension. The effective action, within our two-scalar truncation is:
\begin{equation} \label{7d_2calar_truncation}
\mathcal{S} = \frac1{\kappa_7^2}\int_7 \d^{7}x \sqrt{-g_7}\Bigl( \mathcal{R}-\tfrac{1}{2}(\partial \phi)^2 -\tfrac{1}{2}(\partial\varphi)^2  - \tfrac{1}{2}\Bigl( h\e^{-\tfrac{1}{2}\phi + 6\alpha\varphi} - m\e^{\tfrac{5}{4}\phi+\alpha\varphi}\Bigr)^2\Bigr)\,.
\end{equation}
One can demonstrate that the other scalars in the theory decouple; they are free fields that do not enter the scalar potential (see \cite{Danielsson:2013qfa}). Since these free scalar fields do not enter the superpotential for supersymmetric solutions, they are constant. Hence our truncation is not a restriction for classifying the BPS solutions. Our 7D gauged supergravity has a Minkowski vacuum that breaks all supersymmetry \cite{Blaback:2012mu, Blaback:2013taa}. Hence the effective field theory of fluctuations around the vacuum is not a supergravity. The seven-dimensional gauged supergravity that is obtained from compactification instead captures the spontaneous supersymmetry breaking in the vacuum.  

\subsection{Domain walls from wrapped D8/NS5 branes}

Although the Minkowski vacuum breaks supersymmetry, we do expect 1/2 BPS domain wall flows in the 7D gauged SUGRA. These domain walls should have some description in 10 dimensions as intersecting branes that preserve 1/4 of the SUSY. Such intersecting brane solutions were found by Imamura in \cite{Imamura:2001cr} and the correspondence between the domain walls in 7D and the 10D solutions by Imamura was made in \cite{Blaback:2013taa}. In this paper we elaborate on this correspondence and aim at computing the domain wall tensions in both pictures. 

\begin{figure}[ht!]
	\begin{center}
		\includegraphics{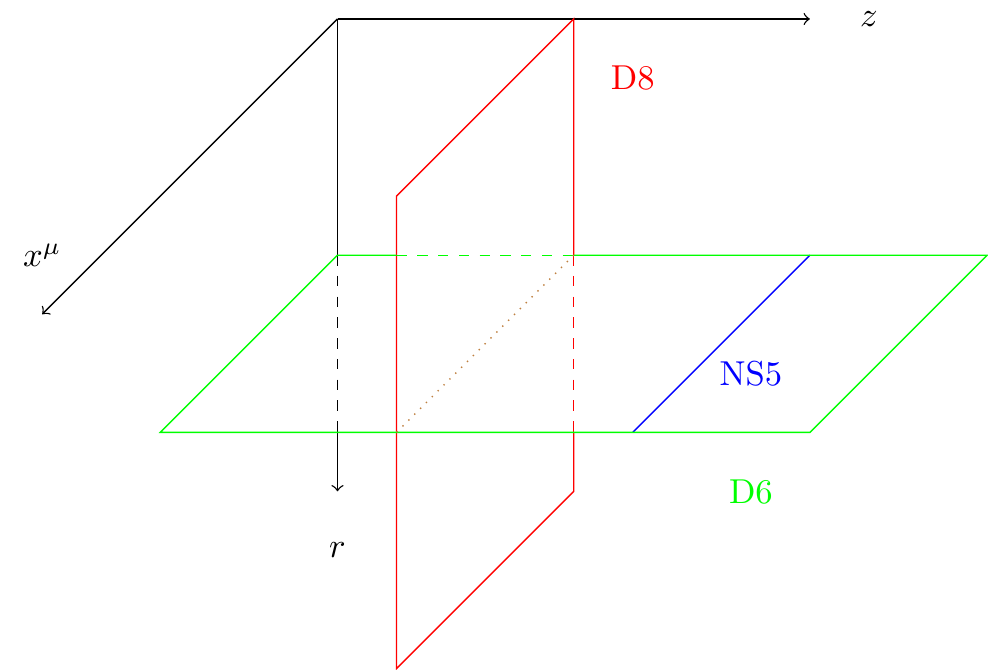}
	\end{center}
	\caption{NS5 and the D8 have one co-dimension inside the D6 worldvolume.\label{fig:partialintersect}}
\end{figure}

The solutions of \cite{Imamura:2001cr}, which we recall in the next section, are reminiscent of D6/D8/NS5 intersections of the form
\begin{align}
\text{D6}:  & \times\,\,|\,\, \times \,\, \times\,\, \times\,\, \times\,\, \times\,\, \times\,\, -\,\,-\,\,- \nonumber\\
\text{D8}: & \times\,\,|\,\, \times \,\, \times\,\, \times\,\, \times\,\, \times\,\, -\,\, \times \,\,\times \,\,\times \nonumber \\
\text{NS5}:  & \times\,\,|\,\, \times \,\, \times\,\, \times\,\, \times\,\, \times\,\, -\,\, -\,\,-\,\,- \label{alignement}
\end{align}
From the intersection diagram it is clear that the solutions are such that the combined effect of the NS5 and D8 intersection is the creation of a co-dimension one object inside the D6 worldvolume as in figure \ref{fig:partialintersect}.

Although these are non-compact D6 solutions, it was argued in \cite{Blaback:2012mu, Blaback:2013taa} that there exists a  formal map between D-brane solutions and O-plane solutions that maps some of the 1/4 BPS solutions of \cite{Imamura:2001cr} to compact O6 solutions with domain walls inside their worldvolume that originate from NS5 and D8 branes. This interpretation was tested in detail by recognizing the field profiles of the domain walls of the lower-dimensional gauged supergravity in the ten-dimensional warpfactors and field expressions of the D6 solutions of \cite{Imamura:2001cr}. It is important to provide more evidence of this correspondence for compact O6 solutions and this is a gap we want to close in this paper by providing the solutions.

We do not expect that all solutions of \cite{Imamura:2001cr} can be interpreted this way using the formal map. For example there is  different class of NS5/D6/D8 intersections that lead to the supersymmetric AdS$_7$ compactification of \cite{Apruzzi:2013yva}. This AdS$_7$ solution can be found by taking the near-horizon limit of the NS5 brane \cite{progress} in a set of solutions of \cite{Imamura:2001cr}.  
  
As a side remark we point out that this set-up can be T-dualised to an intersection of a O(6-$p$)/NS5/D(8-$p$) intersection by T-dualising $p$ directions inside the domain wall living on the D6/O6 \cite{Blaback:2013taa}. For instance for $p=3$ one finds
\begin{align}
\text{O3/D3}:  & \times\,\,| \, \times \,\, \times \,\, -\,\, -\,\, -\,\, \times \,\, -\,\,-\,\,- \nonumber\\
\text{D5}: & \times\,\,|\,\, \times \,\, \times\,\, -\,\, -\,\, -\,\, -\,\, \times \,\,\times \,\,\times \nonumber \\
\text{NS5}:  & \times\,\,|\,\, \times \,\, \times\,\, \times\,\, \times\,\, \times\,\, -\,\, -\,\,-\,\,- \nonumber
\end{align}
This brane intersection can be recognized as the Hanany-Witten set-up \cite{Hanany:1996ie}.
In the case of the O3 compactification the non-compact directions correspond to the O3 worldvolume. The wrapped D5 and NS5 branes have each one co-dimension inside the O3 and they act like domain walls. Therefore the O6 set-up described in this paper can be seen as a simplified version of the general O3 solution with ISD flux.

\subsection{The interpretation of the domain walls} \label{sec:int}
Supergravity domain walls can be classified into two kinds: those that interpolate between two different vacua of the same scalar potential and those that interpolate between vacua of different scalar potentials. The second possibility is more abstract but a simple example is the D8 brane in massive IIA supergravity \cite{Bergshoeff:1995vh, Polchinski:1995df, Bergshoeff:1996ui}. The D8 brane separates regions with different values of the Romans mass $m$: if one passes the D8 wall the discrete jump in $m$ corresponds to the charge of the wall itself. This is manifest in the Bianchi identity for the Romans mass
\be
\d m = Q_8\delta(D8)\,.
\ee
Since the scalar potential of massive IIA supergravity is proportional to $m^2$ we indeed have that the D8 interpolates between regions with a different scalar potential, with the exception of the D8 switching the sign of $m$. This is not any different for the domain walls considered in this paper\footnote{There is an important difference with the D8, since the D8 brane does not have a Minkowsi vacuum at infinity on both sides of the wall. In our set-up we find that the wall separates two Minkowski vacua and this is due to the combined presence of the D8 and the NS5. }. Both at the infinite right and left of the spacetime we will find that the flow reaches the Minkowski vacuum. In order for such flows to stay real, $m$ or the $H_3$-flux quantum $h$ should jump discretely, similar to the D8 brane in 10D space.   So these solutions should be thought of as describing two, possibly coinciding, walls. The position of these walls are determined by the positions in which the quanta $m$ and $h$ make discrete jumps. From a 10D point of view the position at which $m$ ($h$) jumps discretely corresponds to the D8 (NS5) brane. This is in line with the interpretation of the superpotential as the sum of the DBI energies of an D8 and NS5 brane as shown below in equation (\ref{WfromDBI}).

Since our domain walls separate different vacua of 10D string theory with different values for $h$ and $m$, the tadpole condition (\ref{tadpole}) implies that the number of D6 branes differs in each vacua (whereas the number of O6 planes is fixed by the topology of the orbifold). Hence, what effectively happens if one passes through a domain wall is that a certain number of fluxes, carrying D6 charges, is materialised into actual D6 branes or vice versa. The microscopic process that makes this happen can be understood from an open string point of view and proceeds via the motion of D8 and KK5 branes \cite{Gautason:2015tla}. 

These brane-flux transitions will not occur spontaneously since the vacua carry the same energy, at least at the classical supergravity level. However, once supersymmetry is broken by a small amount by some ``uplifting'' effect these domain walls can represent actual vacuum transitions \cite{Ceresole:2006iq, Ahlqvist:2010ki} that appear via the nucleation of spherical bubbles \cite{Brown:1988kg}. These spherical bubbles expand because the energy inside the bubble is smaller than the energy outside. The bubble wall itself corresponds to a spherical D8 or NS5 brane and when these bubbles grow to infinite size they effectively become the stationary infinitely long domain walls considered in this paper, up to the SUSY breaking effects. It has been claimed that the supersymmetry breaking effects can be subleading \cite{Ceresole:2006iq}, although this has been questioned for the case of anti-brane uplifting \cite{Danielsson:2014yga}. In case the supersymmetry breaking effects are indeed negligible then the tension of these domain walls determine the nucleation probability $P$ for the vacuum bubbles according to a standard formula \cite{Brown:1988kg} that reads as follows for 7D theories: 
\be
P\sim \exp\left[-\frac{248832\, \pi^{3}}{35} \frac{T^{7}}{(\Delta \Lambda)^6}\right]\,,
\ee
in natural units, where $T$ is the tension of the walls and $\Delta \Lambda$ is the difference in energy (cosmological constant) between the inside and the outside of the bubble. Hence the tension determines the probability for vacuum decay. The tension can be deduced from the energy $E$ of the domain wall spacetime through the relation
\be
E \Delta t= \int_6 \sqrt{-g_6}\,T\,.
\ee
Where the integral runs over the wall coordinates and $g_6$ is the induced metric on the wall. Since the solutions are static the integral over time gives a formal infinite contribution, $\Delta t$, that can be factored out. 
The energy itself can be obtained by considering the on-shell value of the action, which turns into a total derivative
\be\label{energy}
E \Delta t =\mathcal{S}_{\text{on-shell}}=\int_7 \d(\sqrt{-g_6}W)\,.
\ee
Hence the energy and tension relate to the superpotential evaluated at the boundary. In case of \emph{thick} walls this boundary is only plus and minus infinity (in the z-direction), but in our case we need to include the \emph{thin} walls as well. Then the boundary includes the positions of these branes and asymptotic infinity. We show below that the difference in superpotential around the thin walls exactly equals the on-shell DBI actions of the NS5 and D8 brane in our setup. On the other hand the difference between $W$ at left and right infinity cancels out, because $W$ asymptotes to zero. This implies that the total tension is only given by the thin wall contributions, which is perfectly consistent with our 10D picture in which the domain walls are NS5 and D8 branes and hence the tension should only be given by the fundamental NS5 and D8 tension.
Nonetheless a vanishing contribution to the tension from infinity does not imply that there is no energy associated to the spacetime coming from infinity. It turns out that the $\sqrt{-g_6}$ factor blows up in such a way that it cancels out the vanishing of $W$ at infinity and a non-zero energy is left. This will be made explicit in the next section, where we describe the domain wall solutions in 7D gauged supergravity and from a 10-dimensional point of view. It is exactly this non-zero energy contribution that plays a central role in this paper.

\section{The supergravity solutions} \label{SUGRASOL}
 In this section we first describe the 1/2 BPS domain walls in 7D gauged supergravity, which lift to \emph{smeared} O6 solutions in massive IIA. Then we describe the lift to 10-dimensional solutions with \emph{localised} O6 planes. 
 
\subsection{1/2 BPS domain walls in 7D}
Consider the compactification Ansatz (\ref{oxidation}) that applies to smeared O6 planes. The effective theory is then the gauged supergravity which allows a simple two-scalar truncation (\ref{7d_2calar_truncation}) that captures all 1/2 BPS domain walls.

The 1/2 BPS domain walls are described by a warped metric of the form 
\begin{equation}\label{domainwallAnsatz}
\d s^2_7 = g(z)^{-8}\d z^2 + g(z)^2\eta_{ab}\d x^{a}\d x^{b}\,,
\end{equation}
where $\eta_{ab}$ is the 6D Minkowski metric. The scalar fields only depend on the $z$-coordinate. The BPS solutions can be found from the first-order flow equations
\begin{align}
& \dot{\phi} = -g(z)^{-4}\partial_{\phi} W\,,\qquad
\dot{\varphi} = -g(z)^{-4}\partial_{\varphi} W\,,\qquad \dot{g}(z) = \frac{1}{10} g(z)^{-3}W\,,
\end{align}
with the following real superpotential
\be\label{superpotential}
W(\phi,\varphi) = |h|\e^{-\tfrac{1}{2}\phi + \tfrac{3}{2}\sqrt{\tfrac{3}{5}}\varphi} + |m|\e^{\tfrac{5}{4}\phi+ \tfrac{1}{4}\sqrt{\tfrac{3}{5}}\varphi}\,.
\ee
To anticipate the connection between the domain walls and branes in 10 dimensions (\ref{alignement}) we observe that the superpotential nicely corresponds to the sum of the NS5 and D8 DBI energies
\be\label{WfromDBI}
-\mathcal{S}_{\rm brane} = T_{\textrm{NS}5}e^{-\phi/2}\int_6 \d^{6}x\sqrt{-g_6} + T_{\textrm{D}8}\e^{5\phi/4}\int_9 \d^{9}x \sqrt{-g_9}\,,
\ee
with $T_{\textrm{D}8}=|m|$ and $T_{\textrm{NS}5}=|h|$, where the branes are aligned as in picture (\ref{alignement})\footnote{We do not claim that the wrapped D8 branes and the NS5 branes that act as the domain walls have necessarily the tensions $T_{\textrm{D}8}=|m|$ and $T_{\textrm{NS}5}=|h|$. This would only be true if the vacuum on one side of the domain wall has zero fluxes. In that case the energy of the domain wall equals the superpotential on the other side. When both vacua on the left and the right have non-zero fluxes, the total energy is instead the difference between the superpotentials.}. The fact that the energies sum in the superpotential implies the absence of binding energy as a consequence of the BPS condition.

It is convenient to consider the following base rotation to new scalars $x$ and $u$ \cite{Blaback:2012mu}:
\begin{align}
\phi & = -\tfrac{\sqrt{15}}{8}x +\tfrac{7}{8}u \,,\nonumber\\
\varphi & = - \tfrac{7}{8}x -\tfrac{\sqrt{15}}{8}u\,.
\end{align}
This rotation allows us to embed the superpotential into a class of general superpotential to which the 1/2 BPS domain wall solutions were constructed by Bergshoeff et al.~\cite{Bergshoeff:2004nq}. The warpfactor $g(z)$ can be written as a function of the scalar $x$
\begin{equation} \label{BPS_solution2}
g(z)^{2\sqrt{15}} = e^x .
\end{equation}
Whereas both scalars can be written in terms of two functions $h_1, h_2$
\begin{equation}\label{BPS_solution1}
e^x = (h_1 h_2)^{\tfrac{1}{2}\sqrt{\tfrac{3}{5}}}\,,\qquad e^u = (h_1/h_2)^{1/2}\,,
\end{equation}
where the functions $h_1, h_2$ are linear ``harmonics''
\begin{equation}
h_1 = 2 |h| z + \ell_1^2,\quad h_2 = 2 |m| z + \ell_2^2\,,\label{BPS_solution}
\end{equation}
where $\ell_1^2,\ell_2^2$ are positive integration constants.

Since the domain wall solutions have the property that  at left or right infinity ($z=\mp\infty$) the on-shell potential vanishes
\be
\lim_{z\to \pm \infty} V = 0\,,
\ee
we can think of the solutions as interpolations  between vacua. Within our two-scalar truncation the moduli space of the Minkowski vacua is spanned by the scalar $x$, whereas the scalar $u$ is fixed. The solutions do not flow towards a specific Minkowski vacuum, but instead once the flow reaches the minimum of the potential it flows inside the moduli space such that the scalar $x$ maintains a non-zero velocity. Not only the potential but also the real superpotential obeys
\be \label{Winf}
\lim_{z\to \pm \infty} W = 0\,.
\ee
Note that while $W$ goes to zero, we still have that the density $\sqrt{-g_6} W = g^6 W$ remains finite at the boundary.

The behavior (\ref{Winf}) is naively inconsistent with the `c-theorem' that states that $W$ has to be monotonous along the flow. The way around this are discrete jumps in the parameters $h$ and $m$, which can be thought of as thin walls. Indeed, one can verify along the arguments of \cite{Bergshoeff:1995vh} that the discrete jumps in $h$ and $m$ are needed to find solutions that are well behaved (i.e.~real) over the whole $z$-line. Discrete jumps in $h$ and $m$, should correspond to NS5 and D8 branes respectively, consistent with our interpretation of the superpotential (\ref{WfromDBI}).

 When $m\ell_1^2 = h\ell_2^2$ the scalar $u$ is constant throughout the whole flow, so this special domain wall flows inside the moduli space throughout space. It was the latter domain wall whose lift to massive IIA with localised O6/D6 branes was first realised in \cite{Blaback:2012mu}, since it corresponds to a very simple class of 10-dimensional solutions \cite{Janssen:1999sa, Binetruy:2007tu}. In the next subsection we discuss a generalisation of the solution in \cite{Janssen:1999sa, Binetruy:2007tu} that captures the lift of the 1/2 BPS domain walls with possible different values for $\ell_1^2$ and $\ell_2^2$.

\subsection{1/4 BPS solutions in massive IIA}
The main point of reference \cite{Blaback:2013taa} is that the lift of the 1/2 BPS domain walls to massive IIA with localised sources can be found be rewriting the non-compact solutions of \cite{Imamura:2001cr} in a manner that is independent of the coordinates on the 3-dimensional space transverse to the D6/06 sources. When doing so one can generalize the solutions of \cite{Imamura:2001cr} to compact O6 solutions with domain walls. The solutions take the following form in string frame:
\begin{align}
&\d s^2 = S^{-1/2} \eta_{ab} \d \tilde{x}^a \d \tilde{x}^b + K S^{-1/2} \d\tilde z^2 + K S^{1/2} \d s^2_3\,,\nonumber\\
& \e^\phi = g_s K^{1/2} S^{-3/4}\,,\nonumber\\
&F_2  = -\frac{1}{g_s} \star_3 \d_3 S\,,\nonumber\\
&H_3=  \frac{\partial }{\partial \tilde z} (K S)\star_3 1-  \d\tilde z \wedge \star_3\d _3 K \,, \nonumber\\
& F_0 = m\,,\label{Ima}
\end{align}
where $\d s_3^2$ is the metric on $\mathbb{T}^3/\mathbb{Z}_2$;  $\star_3$ and $\d_3$ are the corresponding hodge star and differential. In local Cartesian coordinates $\theta^i:=y^i/L$ on $\mathbb{T}^3/\mathbb{Z}_2$ we have $\d s_3^2=L^2\,\delta_{ij}\d\theta^i\d\theta^j$. We use tilded coordinates $\tilde{x}^a, \tilde{z}$.
 In the next section (below eq.~(\ref{connectionImasmeared})) we derive a simple relation to the coordinates $z$ and $x^a$ of the previous section. The Ansatz (\ref{Ima}) is given in terms of two functions $S(\tilde z,\theta^i)$ and $K(\tilde z, \theta^i)$ that are determined by the differential equations
\begin{align}
&\nabla_3^2 S + \frac 12 \frac{\partial^2 S^2}{\partial \tilde z^2}  =  -g_s \cQ_6 \delta \,,\label{compactLaplace}\\
&m g_s K = \frac{\partial S}{\partial \tilde z}\,,\label{eq:Imamura_diff_eq2}
\end{align}
with $\delta$ describing the localised O6/D6 sources on $\mathbb{T}^3/\mathbb{Z}_2$ and $\cQ_6:=Q_6/L^3$. Note that this Ansatz is written formally as a D6-NS5 solution. The function $S$ appears with the correct powers for an O6/D6 warpfactor and $K$ for an NS5 warpfactor.

This Ansatz automatically satisfies the tadpole condition \cite{Blaback:2013taa}. Consider the part of the $H_3$ field that sits inside the transversal space:
\begin{equation}
H_3^{\rm internal} =  \frac{1}{mg_s} \frac{1}{2} \frac{\partial^2 }{\partial \tilde z^2} S^2 \star_3 1\,.
\end{equation}
We used the definition of $K$ (\ref{eq:Imamura_diff_eq2}) to rewrite the first term for the expression of $H_3$ in (\ref{Ima}).
Now we rewrite this once more using (\ref{compactLaplace})
such that we end up with
\begin{equation}
H_3^{\rm internal} = \frac{1}{mg_s} (-g_s \cQ_6 \delta - \nabla_3^2 S ) \star_1 \,.
\end{equation}
When we integrate this equation over the compact space the second term on the RHS must vanish and we therefore recover the tadpole condition.

The compactness of the O6 solution implies that the Laplacian-type equation as \eqref{compactLaplace} on a compact space may not be explicitly solvable.  In the non-compact case, where O6 is traded for a D6 and compactness is lost, explicit solutions are known \cite{Imamura:2001cr}. 

The above solutions are 1/4 BPS so they cannot include the Minkowski vacuum Mink$_7\times \mathbb{T}^3/\mathbb{Z}_2$, since that breaks all supersymmetries. The vacuum solution was first found in \cite{Blaback:2010sj} and can be written in terms of above Ansatz by taking $K=1$ and replacing the combination $\partial_{\tilde{z}}(KS)$ in the $H_3$ Ansatz with $mg_s$. The function $S$ then has to obey the following equation in order for all 10D EOM to be solved:
\be\label{Seq}
\nabla^2 S =  g_s \cQ_6 (1-\delta)\,.
\ee
This equation is simpler than the equations describing the domain walls and we analyze the solutions in the appendix of this paper.

\subsection{Smeared limit of the 10D solution}
\label{smeared10d}
The smeared limit is defined as the smearing of the delta-function that describes the O$6$-position, $\delta \to 1$, such that (\ref{compactLaplace}) now has the form
\be
\nabla_3^2 S + \frac 12 \frac{\partial^2 S^2}{\partial \tilde z^2}  =  -g_s \cQ_6  \,,\label{compactLaplace2}
\ee
while (\ref{eq:Imamura_diff_eq2}) remains the same. Consequently, $S$ and $K$ are now only functions of $\tilde z$, meaning the differential equation is simply
\be
\frac 12 \frac{\partial^2 S^2}{\partial \tilde z^2}  =  -g_s \cQ_6  \,.\label{compactLaplace3}
\ee
Since there is no $r$ dependence present, the $H_3$ flux only has one leg
\begin{equation}
H_3 = h \,\d \theta^1 \w \d \theta^2 \w \d \theta^3 := \partial_{\tilde{z}} (KS) \,\d \theta^1 \w \d \theta^2 \w \d \theta^3\,.
\end{equation}

It is then straightforward to derive \cite{Blaback:2013taa} that this 10-dimensional solution with smeared sources correspond to the oxidation of the 7-dimensional solution (\ref{BPS_solution2}, \ref{BPS_solution1}, \ref{BPS_solution}) using the Ansatz (\ref{oxidation}).
To get the expressions to match one needs to rescale coordinates in the proper way by comparing the metric Ansatz of the 7D supergravity \eqref{oxidation} with  10D the solution \eqref{Ima}:
\begin{equation}\label{connectionImasmeared}
h_1 = K^2 S^2 g_s^{-1}, \qquad \text{and,} \qquad h_2 = S^2 g_s^{-2},
\end{equation}
when $\d\tilde{x}^a = g_s^{1/4}  \d x^a$ and $\d \tilde{z} = h_1^{-1/2} g_s^{1/2} \d z$.

Hence we have established that smearing the O6/D6 sources in the Ansatz of Imamura \cite{Imamura:2001cr} then leads to the domain wall solutions of Bergshoeff et.~al.~\cite{Bergshoeff:2004nq} which proves our claim that the solutions of Imamura should contain the domain wall geometries with localised O6 planes.

\subsection{Non-compact solution}
\label{non-compactsolution}
Finding the explicit solutions for the function $S(\tilde{z},\theta_i)$ that describes the domain walls in 7D is most likely an impossible task. In case the $\mathbb{T}^3$ is decompactified to $\mathbb{R}^3$ explicit solutions are easier to find since one can restrict the analysis to solutions with rotational symmetry on the $\mathbb{R}^3$. In practice this means that $S$ only depends on $\tilde{z}$ and $r$, the latter being the radial coordinate on $\mathbb{R}^3$. This was the approach followed in the paper \cite{Blaback:2013taa} that precedes this one and we briefly recall the main idea since it survives when applied to the compact $\mathbb{T}^3$, although explicit expressions are out of reach then. 

It was noted in \cite{Imamura:2001cr} that a radially symmetric solution with O6/D6 sources can be written in terms of a Laurent expansion of $S$ with $1/r$ as the leading singular piece
\be
S(\tilde z, r) = \sum_{n=-1}^\infty b_n(\tilde z) r^n\,.
\ee
The BPS equations then give a \emph{recursive} relation for the coefficients $b_n$ that depend on $\tilde{z}$
\be
n(n+1) b_n = -\frac 12 \frac{\partial^2}{\partial \tilde z^2}\sum_{k=0}^n b_{k-1} b_{n-k-1}\,.\label{eq:sumA}
\ee
This series is solved when the first two terms $b_{-1},b_0$ are given since all the higher order terms are determined recursively.  When $n=0$ the sum \eqref{eq:sumA} leads to $\partial_{\tilde z}^2 b_{-1}=0$ and leaves $b_{0}$ free.  The first term is fixed to be $\tilde{z}$ independent since it has to equal the O6/D6 charge:
\be
b_{-1} = g_s Q_6\,.
\ee
The second term $b_0$ can be constrained by demanding that it corresponds to the smeared solution $S(\tilde{z},r) \rightarrow S(\tilde{z}) = b_0(\tilde{z})$ and hence 
\begin{equation}\label{a_0equation}
\frac 12 \frac{\partial^2}{\partial \tilde{z}^2}b_0^2 = g_s mh\,.
\end{equation}  
By shifting the origin of the $\tilde z$-axis, the most general solution is
\begin{equation} \label{b0mode}
 b_0= \sqrt{g_smh \tilde z^2 + \beta}\,,
\end{equation}
with one integration constant $\beta$ that can be related to the integration constants in the solution of \cite{Bergshoeff:2004nq} as follows
\begin{equation}
\beta = \frac{g_s^2}{h}\left(h\ell_2^2 - m\ell_1^2\right)\,,
\end{equation}
The other terms $b_n$ are now fixed by recursion such that the solution with localised sources is uniquely defined. Other values for $b_{-1}$ and $b_0$ can still solve the equations of motion but cannot be interpreted as the domain walls in 7D.

In the appendix we discuss solutions that break the radial symmetry, by expanding in spherical harmonics.

\subsection{Compact solutions}

For the compact solution on $\mathbb{T}^3/\mathbb{Z}_2$ with orientifold sources, the Laurent expansion should be replaced by a Fourier expansion
\begin{equation}
S = \sum_{\vec{n}}a_{\vec{n}}(\tilde z)\exp[i\vec{n}\cdot \vec{\theta}]\,,
\label{fourierexp}
\end{equation}
where $\vec{\theta}$ are the angles of the 3-torus. 
Both reality of $S$ and the O6 $\mathbb{Z}_2$ involution symmetry $\vec{\theta}\rightarrow -\vec{\theta}$ implies that the coefficients obey:
\be
a_{\vec{n}}^*=a_{-\vec{n}} = a_{\vec{n}} \,.
\ee
If we substitute this into the equation of motion for $S$ \eqref{compactLaplace2}, we find
\be \label{Fourierdecomp}
|\vec{n}|^2\, a_{\vec{n}} = g_s \cQ_6 + \frac{1}{2}\pd_{\tilde z}^2\left(\sum_{m =-\infty}^{\infty} a_{\vec{m}} \, a_{\vec{n}-\vec{m}}\right) 
\ee
for each Fourier mode. 
Unlike the coefficients in the Laurent expansion a  recursive set of equations for the Fourier coefficients $a_{\vec{n}}$ does not exist, instead one finds an infinite set of coupled equations that we could not disentangle. Nonetheless some useful information can be extracted from these equations. In particular, the zero mode must satisfy
\be \label{eqsumans}
\frac12\,\pd^2_{\tilde z}\left(\sum_{\vec{n}} a_{\vec{n}}^2\right) = -g_s \cQ_6\,,
\ee
where we make use of the fact that $a_{-\vec{n}}= a_{\vec{n}}$. Hence
\be \label{sumsoln}
a_0^2 = - g_s\cQ_6\, \tilde z^2 + A\, \tilde z + B\, - \sum_{\vec{n}=1}^\infty a_{\vec{n}}^2, 
\ee
where $A$ and $B$ are integration constants. If we assume that the zero mode of $S$ is uncorrected by the localization of the source, i.e. that it is fixed by the smeared value, we must have that the sum on the RHS is linear in  $\tilde{z}$.

\section{The wall energy} \label{TENSION}
As explained in section \ref{sec:int} the energy of a domain wall relates to the on-shell action (\ref{energy}), which can be written in terms of the real superpotential $W$, within the lower-dimensional supergravity \cite{Cvetic:1996vr}.
We follow a similar reasoning from the 10D point of view and calculate the on-shell action explicitly in appendix \ref{Scomp}. The result is
\be
\boxed{\mathcal{S} = \int_{10} \d^{10}\tilde{x} \partial_{\tilde{z}} \frac{\sqrt{-\tilde g_6}\sqrt{\tilde{g}_{3}}}{g_s^2} \left[\frac{\partial_{\tilde{z}}(KS)}{K} + m g_s K\right].}
\label{10dtension}
\ee
where $\tilde{g}_3$ is the normalised metric on the 3-torus, meaning all factors of $K$ and $S$ have been explicitly accounted for.
The corresponding 7D domain-wall energies can be split into an NS5 and D8 contribution
\be
E_{\textrm{NS5}} = \frac{V_6}{\Delta t} \int_{3} \d^{3}\tilde{x} \frac{\sqrt{\tilde{g}_3}}{g_s^2} \frac{\partial_{\tilde{z}}(KS)}{K}\,,\qquad E_{D8} = \frac{V_6}{\Delta t}\int_{3} \d^{3}\tilde{x} \frac{\sqrt{\tilde{g}_3}}{g_s^2} \left( m g_s K\right) \,.
\ee
Both integrands can be expanded in Fourier modes and the  integral over the compact space picks out the zero-mode only (evaluated at the boundary):
\begin{equation}
E_{\textrm{NS5}} \sim \frac{\partial_{\tilde{z}} (KS)}{K}\Bigg|_0 \qquad \text{and} \qquad E_{D8} \sim K|_0 \,.
\end{equation}
If the zero mode of $S$ is given by the gauged supergravity expression then the D8 contribution to the energy is unaffected since $mg_s K=\partial_{\tilde{z}}S$. The same can be shown for the NS5 contribution but the reasoning is rather involved since the zero mode of $K^{-1}$ does not equal the inverse of the zero mode of $K$.  However, we demonstrate below that at infinite  $\tilde{z}$ the zero mode of $K^{-1}$ becomes the inverse of the zero mode of $K$. Therefore the contribution at infinity stays the same. However, the boundary integral also gets ``thin wall contributions'' at the D8 and NS5 positions exactly match the D8 and NS5 DBI actions as in the smeared set-up.

In the remainder of this section we prove that the zero mode of $K^{-1}$ equals the inverse of the zero-mode of $K$ at infinite  $|\tilde{z}|$:
\be
\lim_{|\tilde{z}|\rightarrow \infty} \Bigl(K^{-1}|_0 - (K|_0)^{-1}\Bigr) =0\,.
\ee
The proof requires that the zero mode of $S$ is given by the gauged SUGRA expression:
\be\label{ao}
a_0(z)=\text{smeared expression for S}.
\ee
In total there are two motivations for (\ref{ao})
\begin{enumerate}
\item Smearing implies deleting the absence on extra-dimensional coordinates and one is left only with $a_0(z)$.
\item The 10D equations of motion do not fix $a_0(z)$. There are an infinite set of solutions and not all can correspond to the domain walls in 7D. Using the energy  as a defining characteristic of a domain wall it is natural to fix the energy for the localised solution by the energy computed in the smeared set-up.
\end{enumerate}
Consider again equation \eqref{eqsumans}. Now recall from the non-compact solution that we have the freedom to pick two modes ($b_{-1},b_0$), while all the other modes follow from the equations. Here we have the same freedom. That means that we can pick $a_0$ to be the smeared result. 
Then from the above equation it follows that
\begin{equation}
\sum_{\vec{n}=1}^\infty a_{\vec{n}}^2 = A\,\tilde z +B \,.
\end{equation}
In particular, we see that the sum grows more slowly than\footnote{This does not preclude any given $a_{\vec{n}}$ from having powers larger than $\tilde{z}$, however. For example, the Taylor expansion of $\sqrt{1+x^2}$ contains all powers of $x$, but it still does not grow faster than $x$.} 
$\tilde z^2$ at large $\tilde z$.

Furthermore, since each $a_{\vec{n}}^2\geq 0$ we are guaranteed that no single term in the sum is greater than the total: they are {\em individually} bounded by
\be
|a_{\vec{n}}| \leq \sqrt{A\,\tilde z+B} \,.
\ee
We then find that, schematically,
\begin{align}
S &= \sqrt{-g_s Q_6 \tilde{z}^2 + \beta} + \sum_{\vec{n}=1}^\infty \mathcal{O}(\tilde{z}^{1/2})\, e^{i\vec{\theta}\cdot \vec{n}}, \label{sassumption}\\
K &= \frac{- Q_6 \tilde{z}}{m\sqrt{-g_s Q_6 \tilde{z}^2 + \beta}} + \sum_{\vec{n}=1}^\infty \mathcal{O}(\tilde{z}^{-1/2})\, e^{i\vec{\theta}\cdot \vec{n}} 
\end{align}
where the zero-mode of $S$ is the same as found in eq.~\pref{b0mode}.

As we now show, these expressions --- which follow from the assumption that the zero-mode of $S$ is unchanged from the smeared result --- predict that the tensions are unaffected by the warping induced by local NS5 and O6 sources. The D8 term straightforwardly agrees with the smeared result, since the integral over the compact space readily selects the zero-mode of $K$. Agreement in the case of the NS5 term in \eqref{10dtension} is somewhat less obvious. We find
\begin{equation}
\partial_{\tilde z}(KS) = \partial_{\tilde z} \left( \frac{-Q_6 \tilde{z}}{m} + \mathcal{O}(\tilde{z}^{1/2}) \right) = \frac{-Q_6}{m} + \mathcal{O}(\tilde{z}^{-1/2}).
\end{equation}
Therefore, for large values of $\tilde{z}$,
\begin{equation}
\frac{\partial_{\tilde z}(KS)}{K} \rightarrow \frac{- Q_6}{m K|_0} + \mathcal{O}(\tilde{z}^{-1/2}),
\end{equation}
which is also equivalent to the smeared result.  The same result is found using the local radial coordinates on the 3-torus as shown in the appendix.

\section{Discussion} \label{Discussion}

Warped compactifications are ubiquitous in string phenomenology, but still warping is insufficiently taken into account in the lower-dimensional effective field theory constructions, because it is not an easy task. In the literature on the topic it is seldom mentioned that ignoring warping is identical to smearing the background orientifold and D-brane sources that sustain the compactification \cite{Grana:2006kf, Blaback:2010sj, Blaback:2012mu}, although this is a clean mathematical way to express the approximation of ignoring warping.

In this paper we have focused on a particular quantity of a warped compactification: the energy of BPS domain walls that interpolate between vacua with different values of the fluxes. We have found that the energy is unaffected by smearing the background branes or, equivalently, by warping.
This is in line with the original arguments that derived the superpotential for flux compactifications based on domain wall tensions \cite{Gukov:1999ya}. 

Our result was derived in two steps, where the separate steps deal with the separate terms in the energy: the contribution from D8 branes and NS5 branes. The D8 part is proportional to the zero mode of the metric function $K$ in the Ansatz (\ref{Ima}). The NS5 part instead is proportional to the zero mode of $K^{-1}$. But these two zero modes do not have to be each others inverse. However, we were able to show that they are in the limit of infinite large $|z|$, which determines the domain wall energy.  It follows that if the zero mode is the same in the warped as in the unwarped case there is no correction. We then argued that exactly the assumption of an unaltered zero mode is what allows one to nail down the 10D localised solution out of a set of infinite solutions to the 1/4 BPS equations in 10D, which we found by extending the results of \cite{Imamura:2001cr}.  Hence instead of finding corrections to the zero mode, the logic is reversed: the energy computed in the smeared limit was the necessary input to find the solution in 10D with all sources localised. The fact that both the NS5 and the D8 energy remained unaffected with the same zero mode is a non-trivial computation that required the knowledge of the 10D 1/4 BPS equations with localised sources.

\subsection*{Acknowledgements}
We would like to thank N.~Bobev, F.F.~Gautason, B.~Janssen, B.~Vercnocke, L.~Martucci, B.~Truijen, M.~Schillo and S. Sethi for useful discussions.
The work of TVR is is supported by the National Science Foundation of Flanders (FWO) grant G.0.E52.14N Odysseus and Pegasus.   The work of JB is supported by the John Templeton Foundation Grant 48222 and the CEA Eurotalents program.  The work of EW is supported by the National Science Foundation of Belgium (FWO) grant G.001.12 Odysseus. MW is supported by a postdoctoral fellowship from the National Science Foundation of Belgium (FWO). We also
acknowledge support from the Belgian Federal Science Policy Office through the Inter-University
Attraction Pole P7/37, from the European Science Foundation through the €˜Holograv€™ Network, and the COST Action MP1210 `The String Theory Universe'.

\newpage
\appendix

\section{10D on-shell action}
\label{Scomp}
In this appendix we compute the explicit 10D on-shell action for the domain walls inside the \emph{localised} O6 planes. As a warm-up we compute the on-shell action for the smeared solution.

\subsection{Smeared limit}

In the smeared limit the on-shell \emph{bulk} action is readily computed to be
\begin{equation}\label{Bulk}
\mathcal{S} = \int_{10} \d^{10}\tilde{x}\,\sqrt{-\tilde{g}_7}\sqrt{\tilde{g}_3} \frac{K^2 }{g_s^2 S} \left[-\frac{\dot{K}^2 S^2}{2K^4} - \frac{h^2}{2 K^4} - \frac{m^2 g_s^2}{2} \right] + \mathcal{B}\,,
\end{equation}
where $\tilde{g}_3$ is the metric on the 3-torus and $\tilde{g}_7$ is the metric on 7D Minkowski space, meaning all factors of $K$ and $S$ have been explicitly accounted for, and $\tilde{g}_{10} =\tilde{g}_7 \tilde{g}_3 $. A dot denotes a derivative with respect to $\tilde{z}$. $\mathcal{B}$ is a boundary term, which cancels against the Gibbons-Hawking term for a constant-$\tilde{z}$ hypersurface. This above expression can be written as a complete-square term as follows
\begin{equation}
\mathcal{S} =- \int_{10} \d^{10}\tilde{x} \,\frac{\sqrt{-\tilde{g}_{10}}}{g_s^2} \left\{ \frac{K^2}{S}\left[ \frac{1}{2}\left( \frac{S \dot{K}}{K^2} - \frac{h}{K^2} + g_s m \right)^2 + \frac{ g_s h m}{K^2} \right] - \partial_{\tilde{z}} \left[ \frac{h}{K} + g_s m K \right] \right\}.\label{smsq}
\end{equation}
If one uses the tadpole condition, $mh=-\cQ_6$, the first term in brackets can be found to be identical to (\ref{compactLaplace3}) and hence vanishes. The second term, proportional to $g_s m h$ drops perfectly against the source term for the O-plane as we now explain. The DBI term is given by
\begin{equation}
\mathcal{S}_{\textrm{DBI}} = -\kappa_7^2\,T_6\int_{10} \d^{10}\tilde{x} e^{-\phi}\sqrt{-g_7} \sqrt{\tilde{g}_3}\,, 
\label{st}
\end{equation}
where $g_7$ is the determinant of the metric on the O6 worldvolume, including all warpfactors. This is the appropriate form of the worldvolume action for a smeared brane. This can most easily be seen by writing the localised action in terms of an integral over 10D space and a localised delta-function. Upon smearing the delta-function becomes $\sqrt{\tilde{g}_3}$.  Since the O6 planes are BPS we must have that $|T_6| = |Q_6|$ and hence $\kappa_7^2 \,T_6 = - hm$, due to the fact that in our conventions $hm>0$ and the O6 planes must have negative tension. Putting everything together the DBI term becomes
\be
\mathcal{S}_{\textrm{DBI}} = h m g_s^{-1} \int_{10} \d^{10}\tilde{x}\, \sqrt{-\tilde{g}_{10}}\, S^{-1}.
\ee
As promised this cancels the $g_s hm$ term in equation \pref{smsq}.

So the final result is
\begin{equation}
\mathcal{S} = \int_{10} \d^{10}\tilde{x}  \frac{\sqrt{-\tilde{g}_{10}}}{g_s^2}  \partial_{\tilde{z}} \left( \frac{h}{K} + g_s m K \right)  =  \frac{1}{\kappa_7^2\,g_s^{2}}\int_7 \d^{7}\tilde{x} \sqrt{-\tilde{g}_7} \partial_{\tilde{z}} \left( \frac{h}{K} + g_s m K \right)\,.
\label{10dsmearedtension}
\end{equation}
In terms of the functions $h_1$ and $h_2$ the energy becomes the following expression evaluated at the boundaries $\partial$:
\begin{equation}
\mathcal{S} = g_s^{-3/2}\Bigl( h(h_2/h_1)^{1/2} + m (h_1/h_2)^{1/2}\Bigr)\text{Vol}_6|_{\partial}\,.
\label{7dtension}
\end{equation}
This expression equals the one found from the gauged SUGRA up to the factor $g_s^{-3/2}$. The extra factor can be explained from the rescaling between the coordinates $x, \tilde{x}$ and $z, \tilde{z}$, as described below (\ref{connectionImasmeared}).
Note that the boundary includes both $z=\pm \infty$ and the thin wall positions. The contributions from the latter equal are accounted for by the usual NS5 and D8 DBI actions. 

The purpose of these domain walls is to connect vacuum solutions. Here, these solutions are Minkowski vacua characterised by the discrete quantities $h$ and $m$. The energy of a domain wall separating a vacuum with $(h_+,m_+)$ from a vacuum with $(h_-,m_-)$ gives
\begin{equation}
\mathcal{S} =  \Bigl(2\sqrt{m_+ h_+} - 2 \sqrt{m_- h_-}\Bigr)\text{Vol}_6.
\end{equation}
This energy is the contribution at $z=\pm \infty$ and not the DBI energies of the NS5 and D8.

\subsection{Localised limit}
We similarly compute the on-shell bulk action for the domain walls inside the \emph{localised} O6 planes. Then both $S$ and $K$ depend on $\tilde{z}$ and the internal coordinates. This computation is rather lengthy as it involves the explicit expression for the Ricci scalar. After eliminating the second derivatives via integration by parts we find
\begin{equation}
\mathcal{S} = \int_{10} \d^{10}\tilde{x} \frac{ \sqrt{\tilde{g}_3}}{g_s^2} \left[-\frac{\dot{K}^2 S}{K^2} - \frac{\dot{K}\dot{S} }{K} - \frac{\dot{S}^2}{2 S} - \frac{m^2 g_s^2 K^2}{2 S} - \frac{(\tilde{\nabla}_3 K)^2}{K^2} - \frac{(\tilde{\nabla}_3S)^2}{ S^2} \right] + \mathcal{B}, \label{brsol1}
\end{equation}
where,
\begin{equation}
\mathcal{B} = -\frac{3}{g_s^2}\int_{10} \d^{10}\tilde{x} \partial_{\tilde{z}} \left[ \sqrt{\tilde{g}_3}\left(\frac{\dot{K} S}{K} - \frac{\dot{S}}{2}\right)  \right] + \frac{1}{g_s^2} \int_{10} \d^{10}\tilde{x} \tilde{\nabla}^2_3 \left[-3 \log K + \frac{5}{2} \log S\right] .
\end{equation}
This first term cancels the Gibbons-Hawking term for a $\tilde{z}$ hypersurface, while the second term is a total derivative on the internal space, so both can be discarded. We now proceed by rewriting \eqref{brsol1} as a sum of equations \eqref{compactLaplace} and \eqref{eq:Imamura_diff_eq2}, and a boundary term. However, as the action also contains derivatives of $K$ with respect to the internal space, we need an additional equation which can be obtained by taking the derivative of \eqref{compactLaplace} and using \eqref{eq:Imamura_diff_eq2}
\begin{equation}
\tilde{\nabla}^2_3 K + \partial_{\tilde{z}}^2(KS)=0.
\label{Kequation}
\end{equation}
The third and fourth term in \eqref{brsol1} can be combined into a square of \eqref{eq:Imamura_diff_eq2}. The extra term introduced to cancel the cross term, can then be combined with the last term in \eqref{brsol1} to form \eqref{compactLaplace}. After some algebra the remaining terms take the form \eqref{Kequation} plus total derivatives. The final result is
\begin{align}
\mathcal{S} = \int_{10} &\d^{10}\tilde{x} \frac{\sqrt{\tilde{g}_3}}{g_s^2 } \Bigg[-\frac{1}{K}\left( \tilde{\nabla}^2_3 K + \partial_{\tilde{z}}^2(KS) \right)- \frac{1}{S}\left(\tilde{\nabla}_3^2 S +mg_s \partial_{\tilde{z}}(KS) + g_sQ\delta  \right) \nonumber \\
&- \frac{1}{2S}\left( \dot{S} - m g_s K \right)^2 +  \frac{g_sQ\delta}{S}  \Bigg] + \int_{10} \d^{10}\tilde{x} \partial_{\tilde{z}} \frac{\sqrt{\tilde{g}_3}}{g_s^2} \left[\frac{\partial_z(KS)}{K} + m g_s K\right] \nonumber \\
& + \frac{1}{g_s^2} \int_{10} \d^{10}\tilde{x} \sqrt{\tilde{g}_3} \tilde{\nabla}_3^2 \left[\log K + \log S \right].
\end{align}
The last boundary term vanishes due to the fact that there are no boundaries in the internal space. The form of the expresison is then the same as in the smeared limit: 
\be
\boxed{\mathcal{S} = \int_{10} \d^{10}\tilde{x} \partial_{\tilde{z}} \frac{\sqrt{\tilde{g}_3}}{g_s^2} \left[\frac{\partial_{\tilde{z}}(KS)}{K} + m g_s K\right].}
\ee
This does not yet show that the actual values are identical.

\section{The vacuum solution}
In this section we consider the vacuum orientifold solution, which contains no domain walls. It's simplicity allows us to infer some general behaviour of these kinds of solutions. For example it was argued in \cite{McOrist:2012yc} that O6 vacua with Romans mass would have problematic backreaction, something we now verify is not the case for our model.

The vacuum solution cannot be found from the BPS equations of \cite{Imamura:2001cr} since it is not SUSY. However its description can be trivially obtained from the Ansatz (\ref{Ima}): take $K=1$ and replace $\partial_{\tilde{z}}(KS)$ in the $H_3$ Ansatz with $m g_s$. Then $S$ obeys
\be\label{Seqapp}
\nabla^2 S =  Qg_s (1-\delta)\,.
\ee
We will solve this equation using Fourier modes as in \eqref{fourierexp}. 
 
The solution is
\be
S = a_0 + Q g_s \sum_{\vec{n}} \frac{1}{|\vec{n}|^2} \exp(i\vec{n}\cdot \vec{\theta})\,,
\ee
where the constant $a_0$ is undetermined. Clearly at the origin, $\vec{\theta}=0$, this solution does not exist since the Fourier series diverges. Most insight can be given by approximating the sum by an integral. This becomes arbitrarily good when sending the size of the compact dimensions to infinity. The result is
\begin{equation}
S = a_0 + \frac{Q g_s}{ \theta},
\end{equation}
where $\theta$ is the size of $\vec{\theta}$. Since $Qg_s$ is negative for O6 planes, the function $S$ is not defined near the origin as expected.  Note that the integral only depends on the size of $\vec{\theta}$, so we get the \emph{radially symmetric solution for S}. Of course we ignored the corrections from long wave lengths, however the effect of their contribution can be guessed from the Euler-Maclaurin formula. But close to the source the S-wave dominates as in flat space solutions.  Indeed in the non-compact limit the continuous Fourier transformation would never see the solutions beyond the s-wave, in agreement with the radially symmetric solutions discussed in \cite{Imamura:2001cr,Blaback:2012mu,Blaback:2013taa}.

A similar result can be obtained when we consider spherical coordinates on the $\mathbb{T}^3$. Such coordinates are only valid locally and it is practically impossible to impose the toroidal boundary conditions in such coordinates. We nevertheless proceed since we would like to show that also in these coordinates locally, around the source, the radially symmetric solution dominates.
The general solution to \eqref{Seqapp} is
\begin{equation}
S = \frac{Q g_s}{r} + b + \frac{Q g_s}{6}r^2 + \sum_{l=1}^\infty \sum_{m=-l}^{l} c_{lm} r^l\, Y_{lm},
\end{equation}
where the $Y_{lm}$ are the spherical harmonics, and the $c_{lm}$ are constants restricted by the toroidal boundary conditions. The first three terms are the vacuum solution found by \cite{Janssen:1999sa} as expected. Again this shows that near the source the spherical solution dominates.

\section{The energy from spherical harmonics}
Here we consider solutions to the differential equation for $S$ \eqref{compactLaplace} using spherical harmonics. The reason is that in this case we are able to find an explicit solution, contrary to the Fourier expansion. However it is not clear that the solution we find satisfies the boundary condition on $\mathbb{T}^3$.

We make the assumption
\begin{equation}
S= \sum_{l=0}^\infty \sum_{m=-l}^{l} r^{-1} U_{lm}(r,\tilde{z}) Y_{lm},
\end{equation}
where the $Y_{lm}$ are the spherical harmonics. With this assumption the equation for $S$ \eqref{compactLaplace} takes the form
\begin{align}
& r^{-1} Y_{00} U''_{00} + g_s Q_6 \delta + r^{-2} Y_{00}^2 \left[\dot{U}^2_{00} + U_{00} \ddot{U}_{00} \right] \nonumber \\
&+ r^{-2} \sum_{l=1}^\infty \sum_{m=-l}^{l} Y_{lm}^2 \left[\dot{U}^2_{lm} + U_{lm} \ddot{U}_{lm} \right] \nonumber \\
&+ r^{-2 }\sum_{l=1}^\infty \sum_{m=-l}^{l} \sum_{l'm' \neq 0,lm} Y_{lm}Y_{l'm'} \left[\dot{U}_{lm} \dot{U}_{l'm'} + \frac{1}{2}U_{lm} \ddot{U}_{l'm'} + \frac{1}{2}U_{l'm'} \ddot{U}_{lm} \right] \nonumber \\
&+ r^{-1} \sum_{l=1}^\infty \sum_{m=-l}^{l} Y_{lm} \left[U''_{lm} - r^{-2} l (l+1) U_{lm} \right] = 0.
\label{spheq}
\end{align}
The first line contains the zero mode and is exactly the same as the non-compact equation in section \ref{non-compactsolution}. If all the other lines are separately zero, this means that the radial symmetric mode is not affected by the higher modes. Integrating equation \eqref{spheq} over the two sphere puts the last two lines to zero, which leaves the second line. The easiest solution puts each term in the sum separately to zero, namely
\begin{equation}
S = s_{wave} + \sum_{l=1}^\infty \sum_{m=-l}^{l} c_{lm} r^{l} \sqrt{2\tilde{z} + c} \, Y_{lm},
\label{sphsol}
\end{equation}
where $s_{wave}$ is the radial symmetric solution to the first line, described in section \ref{non-compactsolution}. The constant $c$ is the same for all modes, but the constants $c_{lm}$ can differ. They will be constraint by the toroidal boundary conditions, but since this is not the most general solution, there is no guarantee that the boundary conditions can be satisfied. However it is striking that the $\tilde{z}$ dependence of the higher modes is the same as for the Fourier solution \eqref{sassumption}.

We now proceed to show that this solution gives no corrections to the energy. Recall that the on-shell action is
\begin{equation}
\mathcal{S} = \int_{10} d^{7}x \,dr\, d\Omega^2_2\, \partial_{\tilde{z}} \frac{r^2 \sin \theta}{g_s^2}\left[\frac{\partial_{\tilde{z}}(KS)}{K} + m g_s K \right].
\end{equation}
From the solution \eqref{sphsol} is is clear that the non-spherical contribution to $K$ goes to zero at $\tilde{z} \rightarrow \infty$ and does not contribute. The same is true for $\partial_{\tilde{z}}(KS)$. To see how the $s_{wave}$ behaves we have to use the solution of section \ref{non-compactsolution}, which is given more explicitly in \cite{Blaback:2013taa}. We find that the following result at $\tilde{z} \rightarrow \infty$:
\begin{equation}
K = \sqrt{\frac{ h }{g_s m}}, \qquad \text{and}, \qquad \partial_{\tilde{z}}(KS) = h.
\end{equation}
The energy is thus exactly equal to the smeared result \eqref{10dsmearedtension}.

\footnotesize{
\bibliography{refs}}

\bibliographystyle{utphysmodb}
\end{document}